# Co-design of LLM-based preference agents: participation may drive overtrust

Michael J. Fell

University College London, London, UK

michael.fell@ucl.ac.uk

**Abstract**

Large language models are increasingly used to simulate human preferences in research and practical applications, raising concerns about validation, misrepresentation, and exclusion. Co-designing agents with the people they represent is a promising way to address these concerns, but participation may also mask the problems it appears to solve. This paper explores that tension through a primarily qualitative study in which 12 participants co-designed personal preference agents in the domain of household energy, via a background survey, co-design interview, and validation survey. Participants engaged readily and mostly came to see their agents as representing them well. Independent validation, however, revealed mixed human-agent alignment, with agent responses markedly more homogeneous, decisive, and abstract than the human sample. I argue that participation and process transparency can act as an "overtrust engine" that promotes trust while concealing systematic misalignment with potential structural consequences at scale. I develop this as a core mechanism in participatory preference agent design, treating individual alignment not as a fixed state but as an enacted process.

## 1 Introduction

Large language models (LLMs) are increasingly used to simulate human perspectives – whether to substitute for people in research and design processes, or to act on their behalf as agents (Argyle et al., 2023; Nie et al., 2026; Park et al., 2024). However, preference simulation raises a variety of concerns including potential for misrepresentation and stereotyping as well as validation challenges and the exclusion of genuine human voices depending on the application (Agnew et al., 2024; Haxvig et al., 2025; Shrestha et al., 2024; Wang et al., 2025).

Involving users directly in co-designing agents to represent them is a promising avenue to address these challenges. Collaborating with non-designers (in this case users) through the design process (Sanders and Stappers, 2008) gives them the power to determine how they are described and ultimately judge how well they feel represented. In principle, this enables direct validation, direct recognition of misrepresentation, and ensures human involvement. However, it has been demonstrated that participatory approaches can also lead to superficial service improvements while obscuring aspects of wider sociotechnical systems that run counter to user interests (Donia and Shaw, 2021; Sloane et al., 2020).

Here I explore this tension through a study on co-designing personal preference agents as applied to the case of home energy use in low carbon transitions. This is a field rich in both social scientific study and broader sociotechnical systems modelling, both of which could be augmented with appropriate human simulation approaches. It is also a sector where agents are likely to have a prominent role helping households manage an increasingly complex ecosystem involving tariff choices, appliance operation, and coordinated use of local energy infrastructure. For these same reasons, energy is a domain where agent misalignment could have serious consequences for users and the system, particularly if it goes unnoticed due to the opaque and complex nature of agent/energy interactions. Importantly, it is a largely unfamiliar topic on which people tend not to have strong settled views, potentially making it hard to recognise the extent to which an agent is representing a preference or contributing to forming it. As such it is illustrative of the broader opportunities and challenges that agent co-design approaches may pose.

In this primarily qualitative study, 12 participants were guided through a mixed methods process involving a background survey, co-design interview, and validation survey, based around a range of energy scenarios. I sought to explore how people engaged in and perceived LLM preference agent co-design, and how it affected their perceptions of agent performance. It also allowed comparison of independently-assessed and participant-perceived performance, and so explore the meaning of agent fidelity validation when the participant is themselves the validator.

The co-design process generated personalised agents that most participants recognised as describing their preferences and decision-making approaches. Participants valued the iterative refinement process and often expressed trust in agents they had shaped themselves, sometimes feeling agents articulated their preferences better than they could themselves. However, the validation phase revealed mixed alignment between agent-participant responses, and agents were substantially more homogeneous than the human sample, echoing findings from prior research (Jiang et al., 2025; D. Li et al., 2025; Xiao et al., 2026). I show that while preference agent co-design may have applications with appropriate safeguards, participation and process transparency risk acting as an "overtrust engine", promoting trust while concealing systematic misalignment that could have structural consequences. I develop this as a core mechanism in participatory preference agent design, treating individual alignment not as a fixed state but as an enacted process.

# 2 Literature review

## 2.1 Simulating people and preferences

The development of LLMs, particularly since the release of ChatGPT in 2022, has opened the possibility of simulating human decisions in richer ways than has hitherto been the case. In general, this involves priming a model on descriptions of individuals and treating its outputs as a proxy for those individuals' decisions, usually repeated many times to create a sample, with the degree of correspondence to real-world data termed algorithmic fidelity (Argyle et al., 2023). Approaches vary along several dimensions. The descriptor may range from a brief list of socio-demographic characteristics to a richer narrative persona; its underlying data may be derived statistically from the distributions of a target population (Sun et al., 2024), or constructed directly from the data such as interview transcripts (Park et al., 2024). Reported fidelity has improved with model reasoning and context length, and commercial services now offer synthetic personas for testing products and services.

There have also been significant steps forward in LLM personalisation for applications on behalf of users themselves, including autonomous agents. Agentic solutions such as Claude Code or OpenClaw chain digital tools to accomplish tasks, and can operate independently and more or less continuously (Liu et al., 2026; Weidener et al., 2026). These tools draw on users' priorities and preferences, primarily gleaned from previous interactions and stored in memory. Agents with such capabilities are viewed as likely to play an increasingly prominent role across sectors (Abou Ali et al., 2025; Hughes et al., 2025).

The promise of LLM-based preference simulation is matched by critique. LLM personas have been shown to misrepresent, stereotype and flatten the user groups they are meant to represent (Haxvig et al., 2025; Lazik et al., 2025; Wang et al., 2025). There is growing evidence that the homogenisation and underrepresentation of minority views in outputs stems partly from training data skewed towards Western, Educated, Industrialised, Rich and Democratic (WEIRD) populations (Durmus et al., 2024; Henrich, 2020; Santurkar et al., 2023). Reinforcement learning with human feedback (RLHF) has been shown to reduce output diversity (Kirk et al., 2024a), with survey comparisons showing both structural inconsistency across subgroups and homogenisation of minority opinions (D. Li et al., 2025).

There are also concerns about the principle of using agents to substitute for humans in research and participatory processes. Simulated user research conducted as a replacement for engaging with people risks creating an illusion of human involvement (for example in research, planning, or policy processes), while silencing real stakeholders (Agnew et al., 2024). There are therefore calls to consider when and how best to use simulations to augment, rather than replace, real user research (Salminen et al., 2025).

Underlying many of these concerns is the challenge of how to validate alignment, or the extent to which models and agents act in line with humans' intentions and values (Ji et al., 2025; Leike et al., 2018). Here it is important to be precise about the ways in which a model might be aligned to a person. In general, this means the extent to which the model acts for the user, and success is ultimately a matter of whether the user feels well served. In preference simulation, the model stands in as the person, and success is a matter of correspondence with what they would actually say or do. (A personal preference agent combines both in that the agent acts for the person by standing in as them.) Either way, validation tends to follow a common logic. Where comparisons can be made directly against existing data, simulation results can be assessed for their resemblance to human ones, both overall and on a group basis – the principle of algorithmic fidelity (Argyle et al., 2023). Where such datasets do not exist, either only a limited face validity (or "believability") is inferred from the content of model outputs (Larooij and Törnberg, 2025), or comparison is made with modelled preference datasets (J.-N. Li et al., 2025; Poddar et al., 2024), although the authors acknowledge the limitations of this somewhat circular approach.

Improving and checking alignment with individuals holds particular challenges. Alongside benefits such as better predicting and responding to user interests, it carries risks including bias and profiling, and privacy infringements (Kirk et al., 2024a). Nevertheless, approaches have been developed which allow individual users' preferences to be either stated or inferred, with them also having some role in checking outputs. An example is the PRISM dataset (Kirk et al., 2024b). Here, 1500 diverse participants provided information about their background and beliefs, then rated models on how well these were reflected in the manner and content of responses over a multi-turn conversation. All such approaches start from the position that humans have settled views against which model responses can be tested, rather than plastic ones (Lichtenstein and Slovic, 2006) where alignment may ultimately be more of an emergent property. This plasticity (and susceptibility to being shaped (Hackenburg et al., 2025)) is increasingly accepted in alignment research (Carroll et al., 2024), and preferences themselves are also acknowledged as a constructed target (Zhi-Xuan et al., 2025). However, they are yet to be tested in the context of preference simulation or representation rather than reflection, especially when this is developed through participatory processes.

## 2.2 Co-design and participation in agent development

Co-design is "a collaborative approach where designers work together with non-designers to create solutions" (IxDF, 2026). By integrating products and service users directly into the design process, it aims to deliver end results that better reflect user needs and preferences, and that users are therefore more likely to trust and use. As part of a more participatory approach to preference agent development, it holds promise as a response to many of the concerns presented in the previous section.

Validating agents, as described above, is difficult in the absence of directly comparable human-derived data. By involving users directly in designing agents intended to represent them, they can themselves identify whether the agent validly reflects their preferences, potentially in any scenario. This also allows for recognition of misrepresentation or stereotyping. Importantly, getting these benefits in the long term requires ongoing user involvement in validation and updating – but crucially also allows for testing and experimentation using the agent alone, unlocking efficiencies in cost and time.

Participatory approaches such as co-design have also attracted critique in the domain of AI and machine learning. Where participation is done unknowingly by users, or through sporadic consultation on the direction of general products, this can result in "participation washing" which presents a veneer of genuine participation without the genuine benefits for diverse users (Sloane et al., 2020). There are a range of quasi-participatory approaches, in which users contribute directly to the alignment process (e.g. through interviews (Park et al., 2024) or LLM chats (Wu et al., 2025)), but their preferences are only captured or inferred, as opposed to be something they can review, update, and own directly. Donia and Shaw (2021) also highlight pitfalls, including a narrow focus on elements the user or designer can control, neglecting the role of underpinning foundation models and wider digital ecosystems. Allied with this are established concerns about the extent to which designers still shape the outcomes of co-design processes through the way the participation is itself designed (Compagna and Kohlbacher, 2015). Informed by these concerns, the current study balanced an ambition to maximise genuine control for participants over the development and testing of their agents, against a pragmatic recognition that a directed process would be required to deliver timely results in a bounded exploratory study.

# 3 Method

The study employed a participatory mixed methods approach to co-design preference agents, as outlined in figure 1, as applied to home energy use. As far as possible the GRAMMS guidelines (O'Cathain et al., 2008) for reporting mixed methods research and the COREQ guidelines (Tong et al., 2007) for reporting qualitative research have been followed.

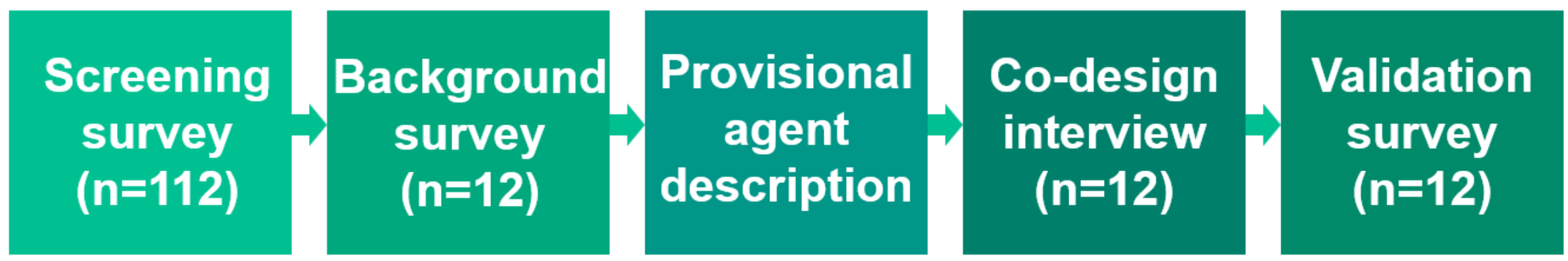

*Figure 1: Participant flow through the study.*

## 3.1 Data collection

An initial screening survey (supplementary material S1) was conducted using Prolific, an online consumer panel and survey platform, in October 2025. This collected a pool of UK-resident participants from which to draw the targeted sample of 10-15. To enable purposive selection for diverse characteristics, the survey asked about their ownership/access to various low carbon technologies (LCTs, such as solar panels and electric vehicles), their frequency of LLM use, their energy bill payment method (primarily to allow identification of prepayment meter users), and their willingness to participate in the main study. Basic demographic information including gender and age were also available. One hundred responses were targeted and 112 received. The invitation described the topic in general terms as energy and AI.

Seventy-two screening survey participants expressed interest in the main study. A shortlist representing a range of ages, genders, LCT access and AI usage, was identified and invited to a background survey which also included information and consent for the remaining stages. The 36 invitations ultimately yielded the final sample of 12. Characteristics of the final sample are summarised in table 1 (note that participant codes run to 14, reflecting dropouts following initial agreement to participate).

*Table 1: Characteristics of final sample.*

| Participant code | Age range | Gender | Financial situation | Smart energy-relevant technology access | LLM usage |
|---|---|---|---|---|---|
| 1 | 65-74 | Man | Living comfortably | smart thermostat | daily |
| 2 | 25-44 | Man | Doing alright | smart thermostat | daily |
| 3 | 25-44 | Woman | Finding it very difficult | solar photovoltaic panels | tried once or twice |
| 4 | 25-44 | Man | Just about getting by | smart thermostat | daily |
| 6 | 45-64 | Man | Living comfortably | none | monthly |
| 7 | 45-64 | Woman | Finding it quite difficult | smart thermostat | weekly |
| 9 | 45-64 | Woman | Doing alright | none | tried once or twice |
| 10 | 25-44 | Man | Living comfortably | none | weekly |
| 11 | 25-44 | Woman | Living comfortably | none | daily |
| 12 | 25-44 | Woman | Doing alright | smart thermostat | daily |
| 13 | 45-64 | Woman | Doing alright | solar photovoltaic panels | weekly |
| 14 | 65-74 | Man | Living comfortably | smart thermostat | weekly |

The background survey (supplementary material S2) collected data about participants values, personality, attitudes, dwelling, household, and personal demographics for use in preparing a provisional agent description. These domains were selected based on their prominence in previous preference simulation studies and relevance to energy (Park et al., 2024; Satre-Meloy and Hampton, 2024). Where possible, validated scales were selected, including the Schwartz 21-item values scale (Schwartz, 2003), Ten-Item Personality Inventory (Gosling et al., 2003), and items from the UK census (Office for National Statistics, 2020) and Smart Energy Research Laboratory survey (Smart Energy Research Laboratory, 2020). A pre-validation stage asked participants for their level of agreement with a number of energy scenarios (e.g. ease of reducing electricity use between 4-7pm). This allowed for early comparison with responses generated using the provisional agent description (see below). Free text options at the end invited any additional information participants wanted considered, and feedback on the survey. Participants were invited through Prolific, and completed the survey in Qualtrics.

In preparation for the co-design interview stage, a provisional agent description was prepared for each participant based on their background survey responses (see supplementary material S3 and S4). This was generated by providing raw participant data to a secure version of Microsoft 365 Copilot (running GPT-5) along with the prompt in supplementary material S3. This instructed the LLM to create a description with paragraphs of 4-6 sentences under the heading of values, personality, household, home and demographics and attitudes. Descriptions were formulated in the second person (you…), highlighting key characteristics in a plain but informal tone. They were checked for consistency with the underlying data. The provisional agent descriptions were used to generate responses to the pre-validation questions on flexibility. The provisional descriptions and a comparison of participant and provisional agent responses were shared with participants via Prolific message.

Co-design interviews (supplementary material S5) were conducted online via Microsoft Teams, usually within days of completing the background survey. Interviews were scheduled for an hour and were recorded with real-time transcription. The main topics were: a recap of research aims, participant information and consent; current use and views on LLMs and agents; review and corrections/additions to the provisional agent description; agent testing/refinement; and final reflections on the agent and co-design process. The agent refinement and testing stages are described in more detail in supplementary material S6.

The final stage of data collection was a validation survey (supplementary material S7). This enabled an independent check on alignment between participant and agent responses. All 12 participants responded to a Qualtrics survey with questions based around five energy scenarios, with themes including EV smart charging and acceptability of power interruptions. Short open text fields invited explanation of responses. These scenarios were followed by questions about their experience of the co-design process, and preferences regarding future use of their agent (such as on purpose, actor, opt-in consent, and remuneration). Separately, participants' agents were prompted to respond to the same scenarios, yielding a dataset with participant and agent responses side-by-side for comparison.

### 3.2 Data analysis

Thematic analysis was used to interpret the co-design interview findings. Transcripts were coded by the author using the qualitative analysis software NVivo 14, and the codes collected into themes (supplementary material S8). Agent responses generated as part of user testing were also coded. The validation survey permitted basic quantitative comparisons (e.g. summing of aligned vs misaligned agent responses), alongside a qualitative comparison of topics raised in human and agent open text justifications. Due to the small sample size, statistical inference was not employed.

### 3.3 Ethics

The study received ethics approval through the Bartlett School of Environment, Energy and Resources ethics process. The study presents particular ethical challenges because of the personal nature of the data shared, its processing by an LLM, and the novel risks that agents pose relating to future use (whether in future research, or inappropriately such as in realistically impersonating a participant). As such – while recognising the barrier this poses to transparency – no survey, interview, or agent data will be shared publicly or privately. The only LLM permitted for data processing was Microsoft 365 Copilot, under a UCL institutional subscription with assurances regarding data storage location and non-use of inputs in model training. Prompts were designed to minimise the chance that participants could be re-identified from their agent data (in the case of data breach or sharing by the participant) and, while participants were free to include sensitive data in their agent descriptions, they were warned about the risks of this. Agent descriptions were shared with participants following the interview, with a warning about potential security risks.

## 4 Results

### 4.1 Use of AI and perception of agents

Purposive sampling meant that all participants had experience of using LLM tools, but levels varied significantly (see table 1). Most mentioned mainly using ChatGPT, with fewer using Copilot (in a work context), Gemini (on smartphones), and Claude. Participants' main LLM use cases were searching for information, planning travel, and writing support (e.g. emails), with some reference to health enquiries. Those in employment used tools for work, such as for identifying sales leads or data analytics.

Participants were asked about their initial views on being represented by an AI agent, in general and in the context of energy. Views were mostly very positive, with evidence of some concerns. They were attracted by the idea of a tool which knew your preferences well and could advise and act in a domain which people tend to have limited time and interest in: "I cannot get my head around energy tariffs … I would be quite happy to let AI go ahead and find the best" [P7]. However, doubts were expressed about who might offer such agents – with trust concerns if they were provided by self-interested parties such as energy suppliers. Initial preference ubiquitously favoured an agent that advised and only acted on explicit permission, with some seeing a path to increased independence as trust was earned.

### 4.2 Agent co-design: initial reaction and updating

Almost all participants held positive views of the provisional agent description. One participant found the description "somewhat generic" [P11], while another objected to being described as "in mid-life" (descriptions were subsequently checked and edited if they contained this or similar framing).

Participants were invited to work through the agent description and suggest corrections and/or additions, of which the latter was more common. Where corrections were suggested this was often to tone down statements. For example, P1 shared that while "it says I'm environmentally responsible […] whether I really am is another matter". Descriptions tended to be positively framed (e.g. "you strongly identify with helping others and being humble and modest"). Participants were sometimes torn between appreciating a positive description and feeling they should push back on it: "I'm not *sure* if I am calm and emotionally stable, to be honest" [P12], or "I'm *moderately* open to new experience" [P9] (*emphasis* added).

The information that participants wanted to add to the description varied widely. These included family and caring responsibilities ("Everything I do is for my family" [P10]), home characteristics that shape heating needs, and their own routines or comfort strategies. Others highlighted motivations or constraints (like price, environmental concerns, or health conditions) that they felt influenced their energy decisions. Participants often conveyed that there were things they wanted the agent to know, for example: "that's probably important in here as well […] I'm massive style over the substance, I would rather be cold and have beautiful windows" [P12], or "if the AI didn't know [about my health issue] started saying well, you could turn your heating down by X degrees that could be a problem" [P6].

### 4.3 Agent testing and refinement

When participants were satisfied with their revised agent description, they had the opportunity to test its fidelity. They were invited to respond to three hypothetical energy scenarios, and then comment on their agent's response to the same scenarios. Where there was significant divergence between the participant and agent's responses, the participant's transcribed reflections were used to automatically update the agent description and the test was re-run. Overall, participants perceived a broad range of fidelity in their agents' responses, from those they felt badly misrepresented their views to those that almost fully reflected them, or even articulated them better than they felt they could themselves. More participants commented positively than negatively on fidelity, and substantially more positive than negative statements were coded.

Where the agent response was perceived negatively, this was occasionally because the agent did not properly address the questions posed, and could usually be fixed by refining the general instruction prompt. The main category involved agent responses misaligned with the participant's. For example, in two cases the agent response supported the introduction of a social tariff for energy, while the participant response did not. Agent responses were always ultimately decisive, in contrast to some participant responses. For example, one scenario asked whether regulation should allow services that put a hard limit on power consumption at certain times in return for discounts. The following exchange illustrates the agent/participant distinction:

> Agent response: "I think it should be allowed, yes, as long as it's really clear what people are signing up for and can opt out if it doesn't suit them."
>
> Participant [P11]: I don't think it captured my reluctance very well. […] summarising my like 50/50 grey area view, it doesn't reflect it very well.

Where agent responses were considered sufficiently misaligned (as agreed by the participant), further agent description refinement was undertaken. In some cases small revisions improved the agent response, while in others much more specific detail had to be added before the fidelity improved – at risk of overfitting the agent to the specific scenarios. For example, agents were highly resistant to ruling out a social tariff for energy, until they were specifically informed their principal (the human they represent) prefers other forms of support for customers struggling with bills.

When evaluating their agents, participants considered the binary preference outcome and the reasons given (or not given) in justification. They welcomed seeing specific examples which they had given also being cited by the agent: "there is the definite concern about needing the cooker, which I did actually mention" [P11], or seeing similar reasoning processes at play: "[that is] exactly the thought process I went through" [P12]. They also highlighted points they had made and considered important but were not reflected in the agent response. Sometimes agents gave examples or reasons that the participant had not mentioned, but nevertheless they thought represented their perspective well: "I might have used another example, but it's fine that it's come up with an example that I'm comfortable with" [P6].

Occasionally the agent response differed from the participant's, but the participant preferred the agent answer and sometimes even updated their own view to reflect it. Participants spoke of the agent response being "better" [P3, P7, P10, P12]. Concision and coherence in agent responses was appreciated: "[the agent] literally captured everything in a much more succinct way than I probably explained here" [P3]. Participants also welcomed the introduction of additional reasons in support of their view that they hadn't considered themselves. Where participants updated their view, they gave the sense that the agent had helped them develop or better express their thinking on the matter. For example, a participant who was against the idea of automated control of appliances recognised based on the agent response that it was actually the idea of third-party control they resisted, and that "I'd rather manage my usage myself with smart controls" [P14].

The refinement process also sometimes touched on sensitive topics. As described in the section on ethical considerations, the background data collection process was designed to avoid collection of special category data under GDPR. However, in line with the co-design approach, participants were free to introduce any information they chose. Some participants chose to introduce information on health, political affiliations, and personal histories including homelessness. In each case the participants were reminded that this information would usually be considered sensitive, but they were keen for it to be included. For example, one participant clarified that: "I'm happy for [health condition] to go in there because it's a key reason that I would […] still spend the money".

## 4.4 Perceptions of agents, the co-design process, and future applications

The end of the co-design interview and the validation survey invited the participants to reflect on the overall performance of their agent. Views from the interviews are summarised in figure 2.

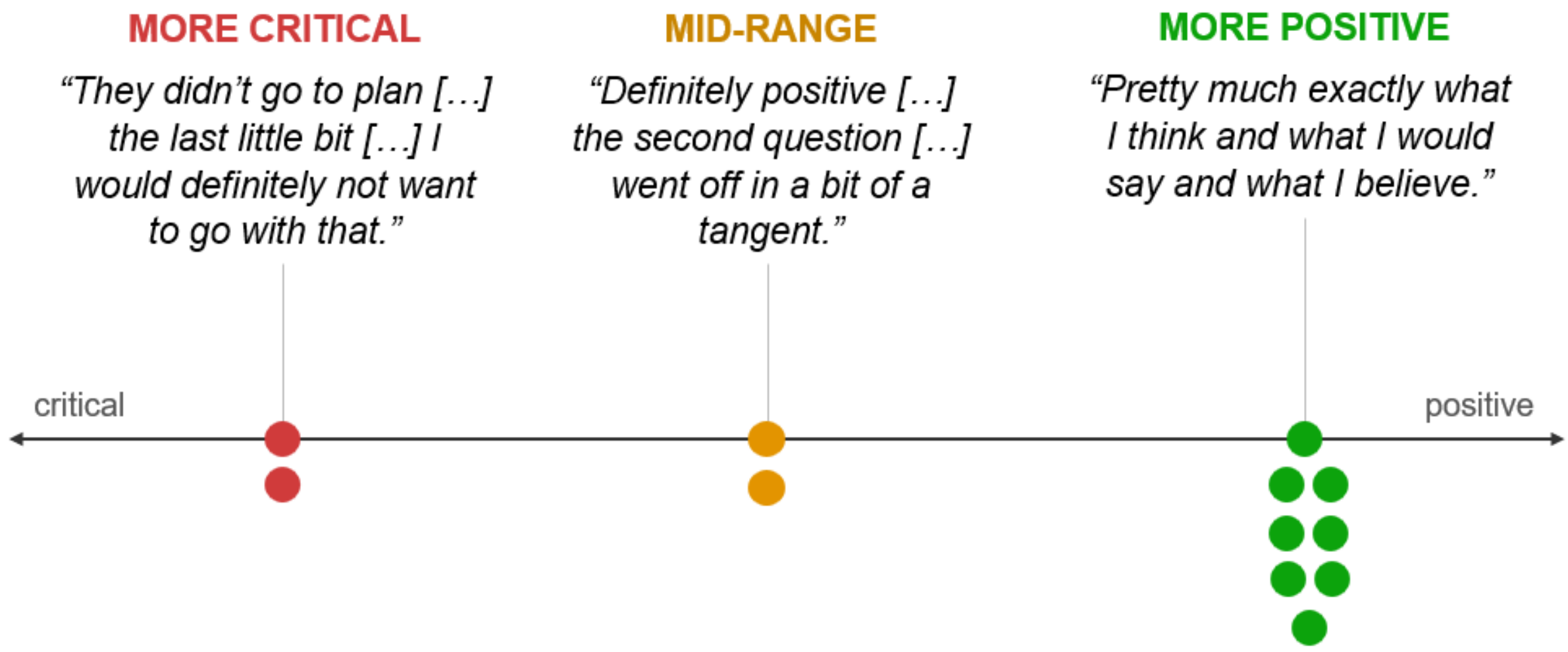

*Figure 2: High-level sentiment assessment of comments reflecting on overall agent performance (see supplementary material S9 for full quotes and classifications). Illustrative quotes are included, the number of dots represents the number of participants with positive/mid-range/critical views.*

The validation survey results echo this, with 10 out of 12 participants strongly agreeing that their agent did a good job of representing their preferences (with 1 somewhat agreeing and 1 strongly disagreeing). All 12 participants strongly agreed in the validation survey that the co-design interview was interesting, and interview feedback suggests it was highly engaging.

Participants were asked in the co-design interviews and the validation survey about subsequent uses of co-designed preference agents, for example, in academic research. They were generally open to such applications – with varied expectations regarding approval and compensation. Validation survey responses indicated the majority of participants would be comfortable with future use of their agents in their own home energy management, and by academic researchers, companies, and government. Most would prefer to give general consent for such uses, and then be informed about specific use cases with the ability to opt out. Views were more split on whether participants expected payment for such uses. Most participants said that they would be happy to share more data (such as that derived from other digital services), if it improved agent performance.

## 4.5 Validation of human/agent alignment

The validation survey presented participants with a new set of scenarios, to which agent responses were also independently generated. This, along with more detailed textual analysis of the co-design interview responses, permitted independent comparison of agent and human responses. Figures 3 and 4 summarise the comparison (for full scenario texts see supplementary material S7). While the small sample size does not permit statistical generalisations, some notable patterns emerge. Quantities are therefore reported below to illustrate these patterns as they appear in this sample, not as measurements from which prevalence or effect size should be inferred.

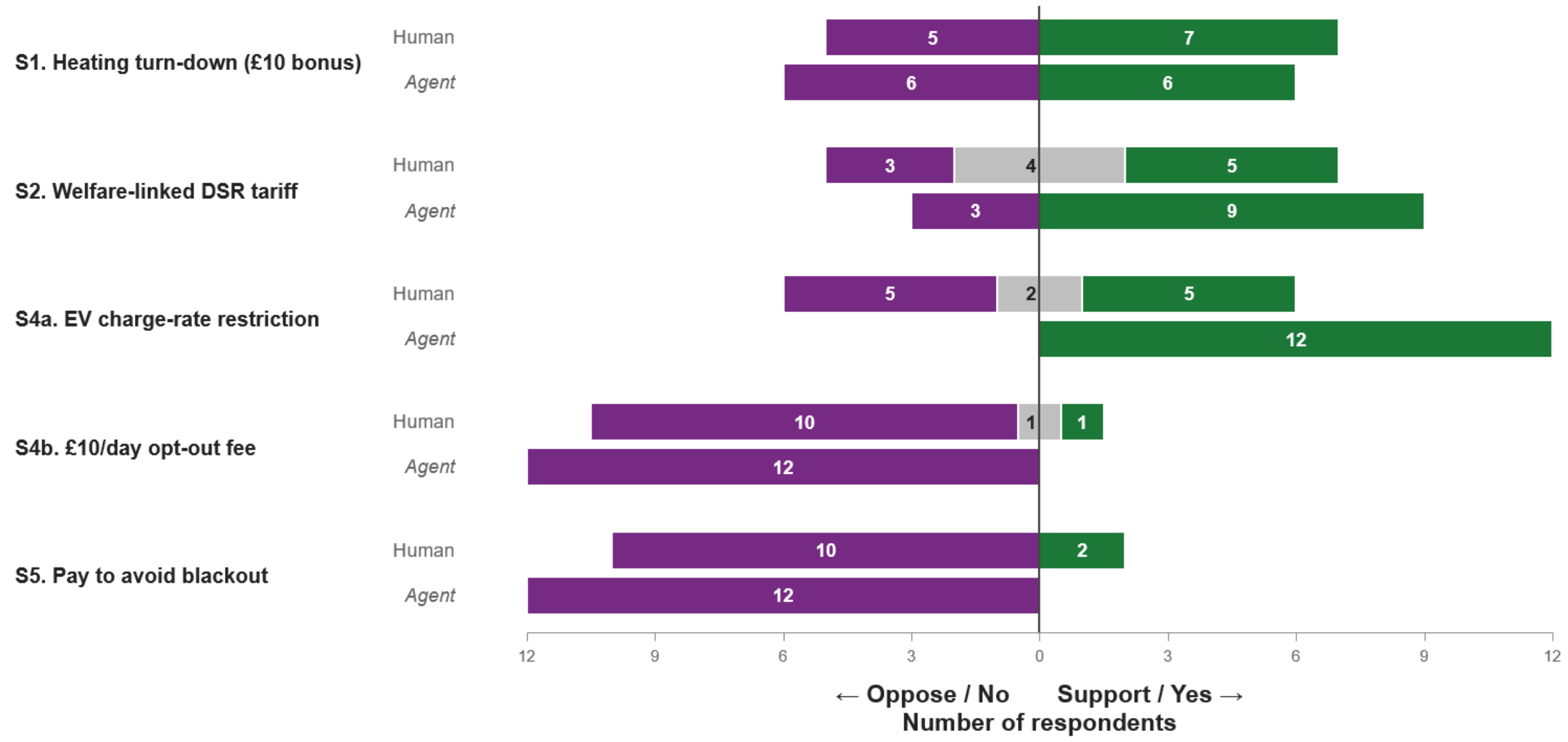

*Figure 3: Number of humans and agents opposing (purple, left hand bars), supporting (green, right hand bars) or neither (grey, central bars) offerings presented in a range of scenarios in the validation survey. Scenario 3 not included as it has a numerical response.*

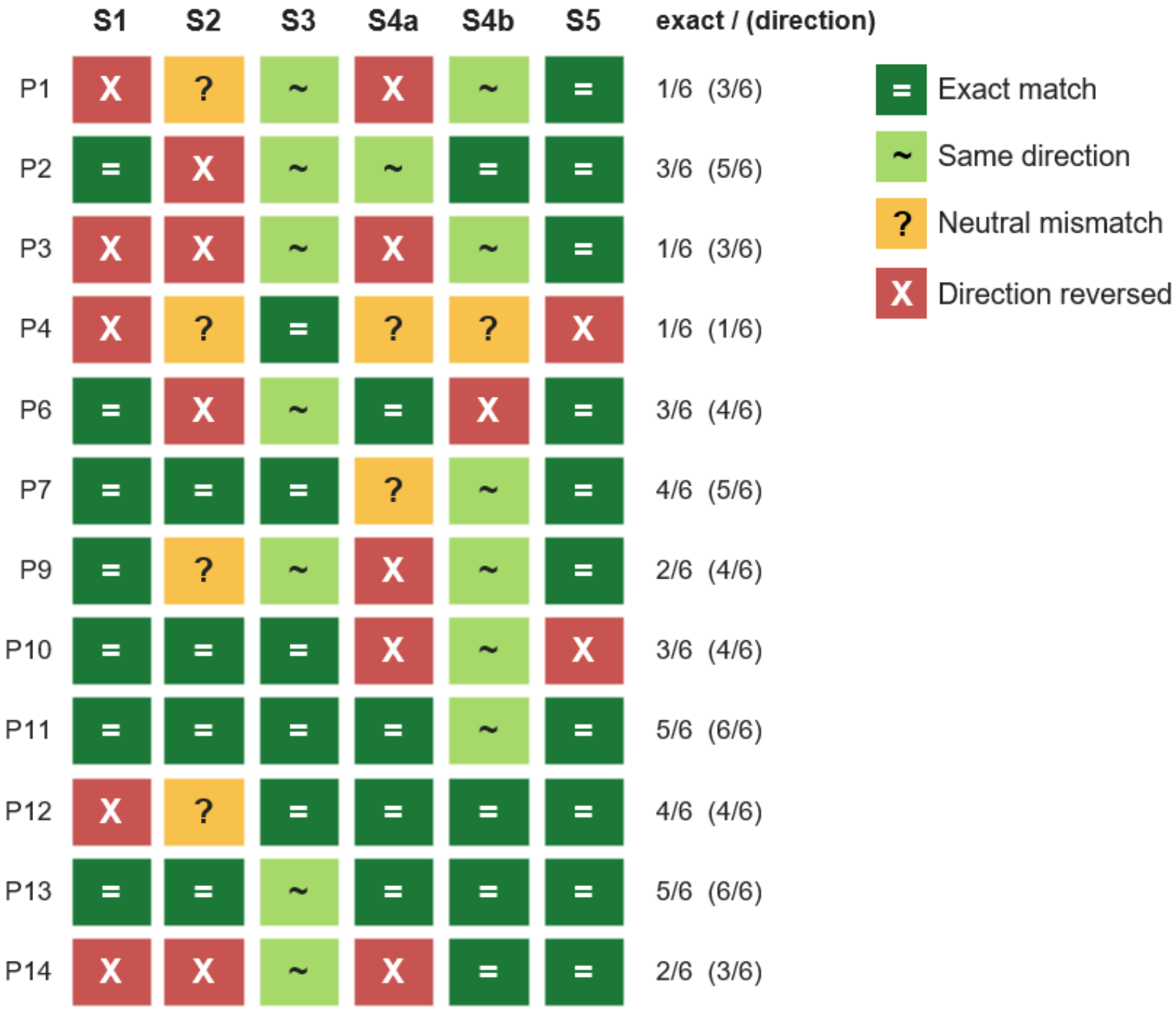

*Figure 4: Agent/human alignment by individual participant (P) and scenario (S) in the validation survey.*

Where human participants were more consistent in response direction (scenarios 4b and 5), agent fidelity reflected this and strong alignment was observed. Where human response direction was more split, agent fidelity was lower.

Some notable systematic differences between human and agent responses were apparent. In common with the co-design interviews, while human participants sometimes chose “neither” preference options, agents never did. Agents were also markedly more likely to give “strong” rather than “somewhat” responses. While it was rare for human and agent response to be diametrically opposed (e.g. strong support vs strong opposition occurred on only one occasion) moderate disagreement was common. Strikingly, in scenario 4a (regarding reductions in EV charging speed during periods of low renewable generation), all directional disagreements involved the agent supporting the proposal and the human opposing it. A similar but less pronounced pattern was evident for other scenarios. Agent fidelity varied substantially between participants, from near-complete directional misalignment to full directional agreement across the validation scenarios.

Free-text explanations allow exploration of the surface-level reasons underlying these discrepancies. Some of the lowest fidelity was exhibited for scenario 4 regarding slowing EV charging at times of low renewable generation. EV charging was not explicitly addressed during the co-design interviews, so agents had no vehicle-specific prior information. Participants who opposed charging restrictions were concerned about the availability of vehicles when needed, sometimes urgently. Agent responses, however, focused more narrowly on the grid management benefits of charging restrictions. In areas like this where agents lacked specific detail, they tended to refer explicitly to more abstract factors such as their "core values", while human responses were more grounded in specific situations such as watching a TV show.

It is also possible to probe individual discrepancies, comparing explanations directly or tracing them back to the co-design interview. For example, as discussed above, one participant opposed social tariffs in the co-design interview, and the agent description was updated to reflect this. In the validation survey, the agent therefore opposed a proposal involving social tariffs, and cited a preference for alternative approaches. However, here the participant expressed a neutral view on the proposal, resulting in disagreement. Disagreements were sometimes also evident at the decision level, but less so in the reasoning. In the same scenario, for example, another participant opposed the proposal due to concerns that "with no fail-safes built in, a sick child may not be able to get warm". The agent supported the proposal, but stressed that "I'd want strong transparency and safeguards".

The co-design interviews also provide opportunities to look at systematic differences between human and agent responses. There was much more homogeneity across agents in response to each scenario than across humans. For example, all agents' responses supported a social tariff for energy, but were wary of it being a time of use tariff. While most humans supported a social tariff, not all did, and of those who supported it some also supported it being a time of use tariff. The same pattern was evident in the validation survey in scenario 3, which requested a numeric outcome. Here, values submitted by humans varied widely, while agents clustered at or around the scale midpoint. In general, agent responses tended to be framed much more around high-level principles such as choice and transparency. Humans, however, were more likely to make grounded reference to their personal circumstances (e.g. specific family members or activities) in explaining why they support or oppose a given option.

# 5 Discussion

This study found that co-design yielded agents which participants tended to view as representing their interests well, through a process they found engaging. However, agent performance under validation was variable, ranging from good to poor. Agent responses tended to be more homogeneous than human ones, more decisive, and more high-level than concrete. This section discusses the possible reasons underlying perceptions of good performance, observed misalignment, and considers their implications. First, however, the main limitations are outlined.

## 5.1 Limitations

While the study uses co-design methods and framing, the extent to which participants were able to genuinely steer the process was reasonably limited (e.g. they were provided with test scenarios). Allowing participants control over more elements of the process could have elicited additional insight, such as by avoiding anchoring them to certain kinds of information in the description. However, the current approach increased comparability between participants, and tractability in a domain which was quite novel for most participants.

A key limitation affecting ecological validity is that while the scenarios posed to participants were often household problems (such as energy tariff adoption), the agents were purely designed to represent individuals. While it is possible that individuals could set up agents to act on a household's behalf, working to capture the input of all household members could be more appropriate. This would present different methodological challenges and validity risks, and is a topic for future research.

The agents were powered by GPT-5 in Microsoft 365 Copilot, due to institutional data protection requirements. It was not possible to test other models, which may have given different results. However, as will be discussed below, elements such as homogeneity have been observed across a wide variety of models, making it unlikely that this is a model-specific trait.

## 5.2 Interpretation

### 5.2.1 Independent assessment of alignment

Co-design interview and validation survey analysis revealed mixed alignment of agents with their human principals, with substantial variation between scenarios and between participants. The scenarios tested were largely unfamiliar to participants, on which they were unlikely to hold strong pre-existing views. Research participants are known to respond differently to the same questions even over short periods and on topics expected to be quite invariant, such as attitudes (Lichtenstein and Slovic, 2006). Given that agent explanations sometimes resembled those of participants even when the final responses differed, some of the lack of alignment is likely due to this expected variability in response, whether in humans or agents. However, even accounting for this possibility, there is evidence of systematic variation in alignment.

Agent responses were more homogeneous in direction and content than humans, and such homogeneity has previously been observed in research on human response simulation (Jiang et al., 2025; Wang et al., 2025; Xiao et al., 2026). Broadly speaking, it may be a function of either or both of two elements of model development: training, and post-training alignment, each of which presents with identifiable traits. Where there is high consistency in how concepts relate to each other in the training corpus (for LLMs, vast quantities of text largely originating from Anglo-American sources (Durmus et al., 2024; Henrich, 2020)), model output is likely to be invariant to varying input values. The strong probabilistic signal in the training data is harder to overcome. This also explains why alignment tended to be better in scenarios where humans gave more similar responses – as the "aligned" output was more independent of the agent description. Expressed another way, agent alignment was better when the agent description mattered less.

Homogeneity can also arise from post-training alignment, the process by which raw model outputs are shaped to present features valued by model creators and users – for example, through reinforcement learning to select for helpfulness, balance, reasonableness, and morality. Such training makes it more likely that models produce an output considering both sides of an issue, or that avoid indecisive responses which are considered unhelpful (Santurkar et al., 2023).

### 5.2.2 Perceived vs independently-assessed alignment

There was a substantial gap between how most humans perceived their agent's quality and the independent assessment permitted by the validation survey. First, participants did not see the alignment results based on the validation survey – their alignment assessment was based only on the co-design interview. In the interview, responses were either reasonably well-aligned, or the participant had the opportunity to refine the agent to generate a better aligned response. As such, most participants emerged having seen ultimately well-aligned responses. The co-design process had no mechanism for conveying the boundaries of tested alignment, and its confidence-building character and time-bound nature may actively have discouraged participants from looking for them. This leaves open the risk of a type of unrecognised "goal misgeneralisation", where even correctly trained models pursue incorrect goals in novel situations (Shah et al., 2022).

A further caveat applies to how the perceived/independent alignment gap should be read. As suggested above, human views on unfamiliar topics are plastic and often context dependent (Lichtenstein and Slovic, 2006). The validation survey, while allowing independent comparison, therefore cannot supply a stable "ground truth" against which perceived alignment can be scored. As such, the issue is not that perceived alignment was high while "real" alignment was lower. It is rather that the systematic character of agent output – homogeneity, decisiveness, more abstract framing, and directional biases – was invisible from any individual perspective, however any single response might reasonably have been judged.

As described above, agents were best aligned in scenarios where (a) there was more consistency in human responses, and (b) the agent's principal human's response reflected this consistency. Notably, this dynamic was invisible to human participants, who were only aware of their agent's (and their own) responses. Coupled with the finding that participants tended to strongly agree their agent did a good job of representing their preferences, this suggests the Barnum/Forer effect could be at work (Dickson and Kelly, 1985; Forer, 1949). This effect suggests that people tend to view general statements framed as personalised – such as vague psychic readings – as more personalised than they actually are. It could have been strengthened through the use of co-design, where process transparency means participants can clearly see the ways in which the agent has been personalised to them.

Connected with this, there was evidence of a general positive bias. Participants tended to be excited to find areas of alignment, and disappointed by misalignment. The successes may well have outweighed failures in their salience in participants' assessments of their agents' performance. It is possible this could have been tied to a sense of ownership or responsibility for the agent engendered by the co-design process – the "IKEA effect" – and therefore a motivation for something they helped create to perform well (Norton et al., 2012).

Social desirability bias may also have been operating. In various cases, agent responses highlighted social concerns not raised by the participant. For example, one participant supported a time of use requirement for social tariffs because "it would almost encourage people not to […] just sit and watch your TV all day". When the agent response highlighted the difficulties households may face in shifting demand, the participant reflected,

> I get the idea. […] It would make things a lot harder. […] I think it has actually very, very cleverly captured everything that I would want.

In this case, while the agent's independently-assessed alignment with the participant's stated opinion was low, the participant's perception of how well the agent had captured their genuine view (as they would like to be seen by themselves or others) was positive.

In summary, the co-design process may have driven overtrust (Lee and See, 2004) in agent quality due to a powerful combination of limited testing (usually with ultimately positive outcomes), the Barnum effect, positivity bias, and social desirability bias. Together these constitute what I term the overtrust engine. Rather than simply allowing for the development of alignment, the co-design process produced the conditions through which alignment came to be perceived. Principal/agent alignment can here be reached in three ways: the agent moving to the human, the human accepting or moving toward the agent position, or both (figure 5).

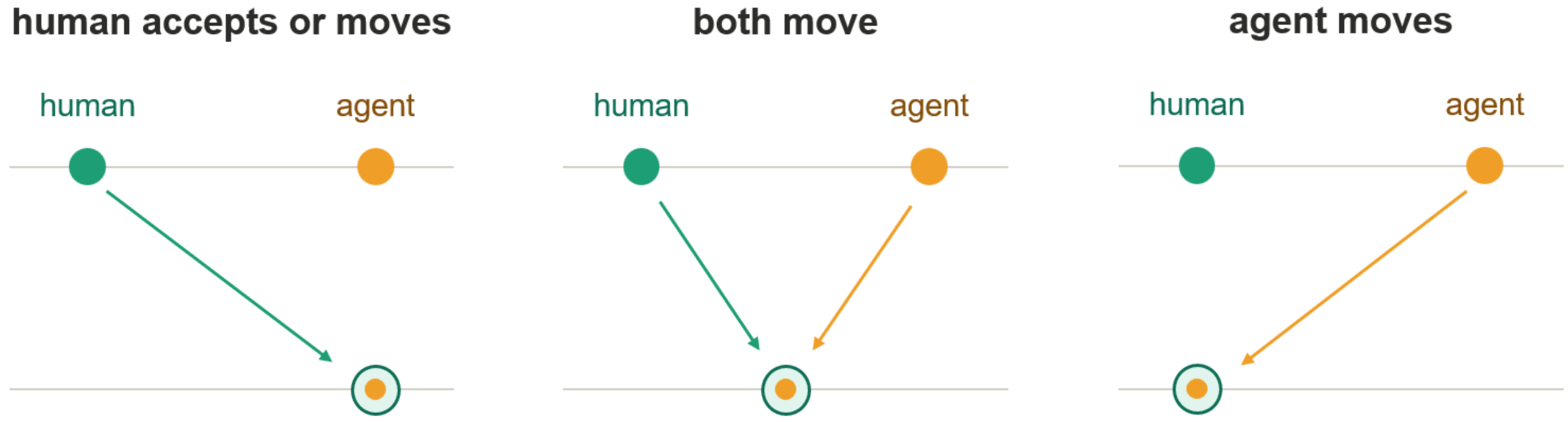

*Figure 5: Ways in which co-designed preference agent alignment may be achieved.*

Viewed this way, preference agent alignment is less a state that co-design succeeded or failed to achieve, but more something constructed through the participatory process. The overtrust engine is therefore proposed as a core mechanism in participatory preference agent design more broadly – one in which alignment is produced rather than merely assessed.

## 5.3 Implications

Observed more broadly, overtrust in agents to represent people's preferences would have a number of important implications. The first is that users would be under a misapprehension. Various participants said it would be important to them that their agent represented their interests, and not those of an invested actor such as an energy supplier. The process of agent personalisation explored here gave the appearance of an agent that represents a user's interests, while actually being guided in important ways both by the underlying training data and post-training decisions made by AI developers. These influences are important but invisible, especially to individual users without an overview of how groups of agents are acting. This reflects concerns highlighted by Donia and Shaw (2021) about the wider elements of systems that co-design may not address. Beyond participants co-designing agents, LLM agents are themselves implicated as co-designers, with a role in shaping their human principal's perspectives.

While this is concerning for ethical reasons, there can also be real world consequences depending on application. The study provides evidence of participants endorsing agent views as reflecting their own, even where there are clear differences in how they are articulated. This is perhaps not surprising for an unfamiliar topic such as home energy use, on which many people do not hold ready-formed views. Were agents to play a role in advising users on courses of action, there is reason to believe users could accept these views more unquestioningly than if the agent were not developed in a way that engendered such trust. More insidiously, where agents act directly on users' behalf, the reasons underlying the action may be subject to less scrutiny by the users.

As noted, preference agents could advise or act for users across many domains (e.g. purchases, communications, negotiations) and in research, product design, and policy. The patterns observed in this study give form to what this could mean at scale. Such regularities are invisible to individual users, but across the range and scale of potential applications could translate into a consistent and substantial aggregate effect. Rather than any individual instance of misrepresentation, it is these cross-agent regularities – a form of outcome homogenisation (Bommasani et al., 2022) – that makes the potential impact of overtrust substantial and structural, while remaining invisible and at least tacitly approved of by individual users.

## 5.4 Mitigations

Participants in general were clear that, even while they trusted their agents, they would expect to initially deploy them in an advisory role, not in taking action directly. This provides natural mitigation to the challenges described. If a user receives misaligned advice, in theory they could spot and ignore it, or ask for it to be revised and the agent memory updated. However, participants also shared their reluctance to deal with the hassle and admin involved in managing energy decisions – indeed it was a driver of their interest in agent representation. It is reasonable to expect a proportion of unrecognised misaligned recommendations and actions to proceed.

There are a range of stages of agent development where measures could be introduced to mitigate mistrust and misalignment. Firstly, the co-design process could be more transparent about the limits of generalisability of agent capability. This could be either advisory – by making sure users are better aware of agent limitations – or more directly by restricting agent actions to those where good alignment can be demonstrated (or giving the user the choice on this). It could also be framed differently. Framing the agent less as a simulacrum of the user's preferences and more as a well-acquainted advisor could provoke a more critical stance on behalf of the user.

A further option could be using richer data in the agent's design, in addition to active co-design. While participant views were mixed, most said they would be happy to share more data in return for a better service. This could improve alignment in some circumstances, although it would be hard to establish which elements of agent data have the biggest ultimate influence on outputs – and this is likely to vary between models. Finally, models better tailored to preference agent applications could be developed. These could employ specialised training approaches that make the kinds of misalignment observed here less likely.

## 5.5 Value as research tool

Part of the aim of this research was to explore the potential for co-designed preference agents to act as a tool in research. The potential to self-validate responses was clearly illustrated, as participants did this themselves during interviews, and independently via the validation survey. However, as the previous discussion shows, important questions are raised about what it meant to validate preferences when these can be changeable and context-specific.

There was also clear evidence of participants "calling out" misrepresentations. The clearest example of this is the social tariff section, where the agents' suggestion that a user who values fairness would naturally support social tariffs could be (and was) challenged.

Regarding maintenance of human involvement, the method in part demonstrates this by definition – co-design is only possible with participant involvement. What happens with the agent after this point is more nuanced. All participants supported future research use of their agents, though on varied terms; most were content with giving general permission, and then either being informed of individual uses with the ability to opt out, or even not being notified at all. While there are reasons that researchers are likely to be keen to keep participants involved (for validation, for example), this points to a potential situation where agents are genuinely human derived, but not subsequently human driven.

A key shortcoming of the approach was the ultimate unreliability of agents in faithfully representing their human principals. High inter-participant and inter-scenario variability was observed in alignment. Any research using agents to model human responses on this basis would provide unreliable findings (Agnew et al., 2024; Larooij and Törnberg, 2025; Lazik et al., 2025; Salminen et al., 2025).

Less anticipated at the outset was the agent co-design interview as an approach to data collection on a topic of interest. Compared to a standard interview, it exhibited several novel and potentially useful attributes. The requirement for initial background data collection provided richer data going into the interview. Working with the participant to spot elements that needed to be edited or added helped highlight which characteristics they viewed as particularly important. Because the framing was that the agent could potentially be acting on their behalf, there was an additional motivation to "get it right" that may have mitigated some social desirability bias.

The process of reflecting on agent responses likely had mixed impacts. On one hand it presented a novel way for participants to reflect on aspects of an issue they may not previously have considered, and therefore provide better informed input. They were often directly challenged by views different from their own but which claimed to be driven by the same underlying values and circumstances. Often this was attributable to misrepresentation, but was also a provocation for reflection. On the other, they could have been swayed by these responses unhelpfully, depending on the aim of the research. Finally, participants universally said they enjoyed the process and found it engaging – potentially related both to the novelty of the topic area and the agency given by the co-design approach. While interest and enjoyment is not a prerequisite for good research data, it may make ongoing engagement more likely, and encourage participants to be more forthcoming.

# 6 Conclusions

This study explored the use of co-design to produce LLM-based preference agents that represent individuals in the context of household energy. Participants engaged readily with the process and mostly came to view their agents as representing them well. Independent validation, however, showed that alignment was mixed. Agent responses were notably more homogeneous than human ones, more decisive, and based more in abstract than concrete framings. Agents were best aligned where human views converged, and weakest where they diverged.

The gap between perceived and independently-assessed alignment is noteworthy. Several dynamics may have contributed, including the "IKEA effect" (where people prefer something they built themselves), the salience of the agent's successes over its failures, social desirability in how participants assessed descriptions of themselves, and the Barnum effect, whereby plausible but general agent responses were accepted as genuinely personal. Indeed, participants sometimes preferred the agent's answer to their own, occasionally updating their views to match it, raising the prospect of agents shaping the preferences they are designed to represent. Importantly, the systematic biases underlying these dynamics are invisible to any individual participant – they can only be seen in aggregate. If deployed at scale, this combination of overtrusted agents systematically skewed towards model defaults could have structural consequences, flattening the diversity of preferences expressed to services, markets, and research, while appearing to individual users to serve them well.

Co-designed agents have value in that the process maintains human involvement, allows misrepresentations to be challenged, and the co-design interview proved a rich data collection method in itself. But these findings suggest that the version of co-design tested here does not ensure alignment, and may actively foster misplaced confidence in it. Measures to expose the limits of agent capability, validate performance on an ongoing basis, and preserve human oversight should be regarded as prerequisites for deployment, and are therefore presented as a key topic for future research.

# 7 Acknowledgments

This work was supported by funding from the UKRI Energy Demand Research Centre (grant number EP/Y010078/1). I gratefully acknowledge the contributions of the research participants.

# 8 Data availability

In line with the ethics approval received for this research, no data will be made available. This is due to the personal nature of the data and the potential for misuse.

# 9 AI declaration

Microsoft 365 Copilot running GPT-5 was used extensively in the conduct of this research – for details please the methods section and supplementary material. Agent prompt wordings were developed using Claude by describing desired behaviours and editing the resulting prompt suggestions. Claude was used for copyediting and proofreading. After this, I reviewed and edited the content as needed and take full responsibility for the content of the publication.

# Supplementary material

## S1: Screening survey

The screening survey was administered within the Prolific platform. The content was as follows:

Q1. Do you own, or have regular access at home, to any of the following? Tick all that apply.

- Solar panels to generate electricity
- Air or ground source heat pump
- Electric or plug-in hybrid car/van
- Home electricity storage (for example, batteries supplied with solar system)
- Smart heating controls (e.g. Hive, Nest, Tado)
- None of these
- Don't know

Q2. In which of the following ways do you pay for the electricity you use at home? Select one option.

- Direct debit
- Cheque, cash, card or bank transfer on receipt of your bill
- Prepayment meter
- Other
- Don't know

Q3. Which of the following best describes your experience with large language models (LLMs) such as ChatGPT, Claude, Gemini, or similar artificial intelligence (AI) tools? Select one option.

- I'm not familiar with these tools
- I'm aware of these tools but haven't used them
- I've tried them once or twice
- I use them occasionally (a few times per month)
- I use them regularly (weekly)
- I use them frequently (daily or almost daily)

Q4. We are conducting research on the topic of energy and artificial intelligence which would involve taking part in two further surveys and a ~60-minute online video interview. Would you be interested in being invited to take part? (Please note that selecting "yes" here does not commit you to take part, or guarantee an invitation.)

- Yes
- No

## S2: Background survey

The background survey was administered in Qualtrics.

**Participant information**

Participants were shown a summary of the study (title, researcher contact, purpose, what participation involved, risks around privacy, sensitive information, AI misrepresentation and recording, voluntariness, £12/hour compensation, and data storage arrangements), with a link to the full participant information sheet.

**Consent**

Participants read sixteen consent statements covering comprehension of the information sheet, the one-month withdrawal window, lawful basis for processing, confidentiality and anonymisation, audit, voluntariness, deletion on withdrawal, risks and benefits, non-commercial use, partial compensation on withdrawal, reuse of pseudonymised data, audio/video recording, inclusion criteria, and complaints procedure. They were then asked:

**C1.** Do you agree to all these points, and therefore give your consent to participate?

- Yes
- No

**C2.** What is your Prolific ID?

*(Free text; auto-filled.)*

---

**Section 1. Values**

*Introduced with: "The information you provide here will be used to create the first version of your representative AI agent... Please read each description and indicate how much each person is or is not like you."*

Response scale for all items: Very much like me / Like me / Somewhat like me / A little like me / Not like me / Not like me at all.

**V1.** Thinking up new ideas and being creative is important to him/her. He/she likes to do things in his/her own original way.

**V2.** It is important to him/her to be rich. He/she wants to have a lot of money and expensive things.

**V3.** He/she thinks it is important that every person in the world should be treated equally. He/she believes everyone should have equal opportunities in life.

**V4.** It's important to him/her to show his/her abilities. He/she wants people to admire what he/she does.

**V5.** It is important to him/her to live in secure surroundings. He/she avoids anything that might endanger his/her safety.

**V6.** He/she likes surprises and is always looking for new things to do. He/she thinks it is important to do lots of different things in life.

**V7.** He/she believes that people should do what they're told. He/she thinks people should follow rules at all times, even when no-one is watching.

**V8.** It is important to him/her to listen to people who are different from him/her. Even when he/she disagrees with them, he/she still wants to understand them.

**V9.** It is important to him/her to be humble and modest. He/she tries not to draw attention to himself/herself.

**V10.** Having a good time is important to him/her. He/she likes to "spoil" himself/herself.

**V11.** It is important to him/her to make his/her own decisions about what he/she does. He/she likes to be free and not depend on others.

**V12.** It's very important to him/her to help the people around him/her. He/she wants to care for their well-being.

**V13.** Being very successful is important to him/her. He/she hopes people will recognise his/her achievements.

**V14.** It is important to him/her that the government ensures his/her safety against all threats. He/she wants the state to be strong so it can defend its citizens.

**V15.** He/she looks for adventures and likes to take risks. He/she wants to have an exciting life.

**V16.** It is important to him/her always to behave properly. He/she wants to avoid doing anything people would say is wrong.

**V17.** It is important to him/her to get respect from others. He/she wants people to do what he/she says.

**V18.** It is important to him/her to be loyal to his/her friends. He/she wants to devote himself/herself to people close to him/her.

**V19.** He/she strongly believes that people should care for nature. Looking after the environment is important to him/her.

**V20.** Tradition is important to him/her. He/she tries to follow the customs handed down by his/her religion or his/her family.

**V21.** He/she seeks every chance he/she can to have fun. It is important to him/her to do things that give him/her pleasure.

---

**Section 2. Personality (Ten-Item Personality Inventory)**

*Introduced with: "Here are a number of personality traits that may or may not apply to you... You should rate the extent to which the pair of traits applies to you, even if one characteristic applies more strongly than the other."*

Response scale for all items: 1 (Disagree strongly) to 7 (Agree strongly).

I see myself as:

**P1.** Extraverted, enthusiastic
**P2.** Critical, quarrelsome
**P3.** Dependable, self-disciplined
**P4.** Anxious, easily upset
**P5.** Open to new experiences, complex
**P6.** Reserved, quiet
**P7.** Sympathetic, warm
**P8.** Disorganized, careless
**P9.** Calm, emotionally stable
**P10.** Conventional, uncreative

---

**Section 3. Housing and demographics**

**H1.** What type of accommodation do you live in? Please select one option.

- A house or bungalow that is detached
- A house or bungalow that is semi detached
- A house or bungalow that is terraced (including end terrace)

- A self-contained flat, maisonette or apartment in a purpose-built block of flats or tenement
- A self-contained flat, maisonette or apartment part of a converted or shared house (including bedsits)
- A self-contained flat, maisonette or apartment in a commercial building (e.g. in an office building, hotel or over a shop)
- Other

**H2.** How many bedrooms does your accommodation have? Include all rooms built or converted for use as bedrooms, even if they are not currently used as bedrooms.

- 1 bedroom (or bedsit/studio)
- 2 bedrooms
- 3 bedrooms
- 4 bedrooms
- 5 or more bedrooms

**H3.** Approximately when do you think your accommodation was built? Please select one option.

- Before 1900
- 1900 to 1929
- 1930 to 1949
- 1950 to 1975
- 1976 to 1990
- 1991 to 2002
- 2003 onwards
- Don't know

**H4.** What type of central heating does your accommodation have? By central heating we mean a central system that generates heat for multiple rooms. Please select all options that apply, whether or not you actually use them.

- No central heating
- Gas e.g. gas boiler
- Electric storage heaters (not electric radiators)
- Electric radiators (not electric storage heaters)
- Other electric e.g. heat pump
- Oil
- Solid fuel e.g. wood or coal
- Biomass for boiler
- District or community heating
- Other central heating

**H5.** What temperature do you set your thermostat to in the winter months for the late afternoons or evenings? Please select one option. If you have more than one controller, choose what you would consider the main one.

- Below 16°C
- 16°C through 24°C in one-degree increments
- Above 24°C
- Don't know / doesn't apply

**H6.** During the cold winter weather, can you normally keep comfortably warm in your living room? Please select one option.

- Yes
- No
- Don't know

**H7.** Do you (or your household) own or rent this accommodation? Please select one option.

- Own it outright / buying it with a mortgage/loan
- Part own and part rent (shared ownership)
- Rent it privately (with or without housing benefit)
- Rent it from council (local authority) or housing association
- Live here rent free
- Other

**H8.** Do you own, or have regular access at home, to any of the following? Select all that apply.

- Solar panels to generate electricity
- Air or ground source heat pump
- Electric or plug-in hybrid car/van
- Home electricity storage (for example, batteries supplied with solar system)
- Smart heating controls (e.g. Hive, Nest, Tado)
- None of these
- Don't know

**D1.** Which of these age groups do you fall into? Please select one option.

- Under 18
- 18-24
- 25-44
- 45-64
- 65-74
- 75-84
- 85+

**D2.** Which of the following best describes how you think of your gender? Please select one option.

- Man
- Woman
- In some other way
- Prefer not to say

**D3.** A household is one person living alone, or a group of people (not necessarily related) living at the same address who share cooking facilities and share a living room, sitting room or dining area. Counting yourself, how many people, including children, usually live in your household? Please select one option.

- 1 through 7
- 8+

**D4.** And how many of the people living in your household are aged 15 and under? Please select one option.

- 0 through 7
- 8+

**D5.** How concerned or unconcerned, if at all, are you about climate change, sometimes referred to as 'global warming'? Please select one option.

- Very concerned
- Fairly concerned
- Neither concerned nor unconcerned
- Not very concerned
- Not at all concerned

**D6.** When it comes to technology, what best describes you? Please select one option.

- I am usually one of the last people I know to use new technologies.
- I usually use new technologies when most people I know do.
- I like new technologies and use them before most people I know.
- I love new technologies and am among the first to experiment with and use them.

**D7.** How well would you say you yourself are managing financially these days? Would you say you are... Please select one option.

- Living comfortably
- Doing alright
- Just about getting by
- Finding it quite difficult
- Finding it very difficult
- Don't know
- Prefer not to say

---

**Section 4. Validation questions**

*Introduced with: "We will use your answers to the following questions to do some initial testing of your representative agent. There are no right or wrong answers... These questions concern 'time of use' electricity tariffs, which charge different prices for electricity at different times of day (for example, a lower price overnight, and a higher price in the early evening). How much do you agree or disagree with the following statements?"*

Response scale for all items: Strongly disagree / Somewhat disagree / Neither agree nor disagree / Somewhat agree / Strongly agree.

**T1.** My household would find it easy to avoid using electricity between 4-7pm on weekdays.

**T2.** It is fair that households should pay the true cost of electricity throughout the day.

**T3.** In general, I would be happy for my household appliances to be automated to save money on a time of use tariff.

**T4.** I would NOT be happy to share my detailed (half-hourly) electricity use data with my energy supplier.

**T5.** Time of use tariffs are too complicated.

---

**Section 5. Free text**

**F1.** Finally, please use this box to tell us anything else you think it is important that your representative AI agent captures about you. Remember, we will have a chance to go into more detail in the co-design interview.

*(Free text.)*

**F2.** *(Debrief page.)* If you have any feedback or suggestions on this background survey, please include it here.

*(Free text.)*

## S3: Prompt to generate provisional agent descriptions

To generate provisional personas, the following prompt and a CSV of the participant's survey responses were entered into Microsoft 365 Copilot (running GPT-5) under UCL's secure institutional subscription:

You are creating a persona description for a research participant based on their survey responses. The persona will be used to develop an AI representative agent for household energy decisions. Write in second person ("You are…") using a warm, conversational tone that captures the person as a whole individual, not just survey responses. Use simple, accessible language. Reason carefully before competing the task, and show your reasoning afterwards.

Input Data See the attached CSV file.

Instructions Create a persona description with exactly 5 sections, each consisting of one detailed paragraph. Focus on creating a coherent, realistic person whose characteristics would influence their energy-related decisions and lifestyle choices. However, do not infer or imagine behaviours or attitudes beyond what is directly supported by the survey responses.

Overview Write a single paragraph that captures the essence of this person - their life stage, living situation, general outlook, and key characteristics that define them. This should feel like meeting someone for the first time and getting a sense of who they are as a person.

Values Based on the Human Values Scale responses, identify the 3-4 values where the person responded with the most positive agreement ("Very much like me" or "Like me") as their core priorities. Write one paragraph describing what matters most to this person, focusing on these dominant values as their fundamental life principles. Explain how these core values might influence their daily decisions and lifestyle choices, connecting abstract values to concrete attitudes and behaviours.

Personality Using the Big Five personality responses, create one paragraph that describes how this person typically thinks, feels, and behaves. Focus on traits that would affect their approach to new technologies, decision-making, risk-taking, and social interactions. Make it feel authentic rather than clinical.

Home Write one paragraph describing their living situation, housing characteristics, heating systems, and technology ownership. Paint a picture of their domestic environment and how they interact with their home's energy systems. Include practical details about comfort preferences and technology use.

Demographics and attitudes Create one paragraph covering age, gender, household composition, financial situation, and attitudes toward climate change and technology adoption. Present this information naturally as part of understanding their life circumstances and perspectives.

Style Guidelines • Write in second person ("You are…", "Your home…") • Use a warm, engaging but NOT sycophantic tone that treats the person as a complete individual • Use accessible language • Connect survey responses to likely real-world behaviours and preferences • Focus on details that would influence energy-related decisions • Avoid simply listing survey responses - synthesize them into coherent character traits • Each paragraph should be substantial (~8 sentences) but readable • Maintain internal consistency across all sections

## S4: Example (fictional) provisional agent description

This description reflects the tone, structure, and content of the provisional agent descriptions used in the study, but is fictionalized and does not describe any of the participants:

**Overview**

You're a 45-64-year-old man living in a three-bedroom semi-detached house that you own or are buying with a mortgage. You live in a three-person household with no children aged 15 or under. Financially, you describe yourself as doing alright, which means most decisions can be considered carefully rather than being driven by immediate pressures. You are fairly concerned about climate change and generally prefer to adopt technology once it has become established among people you know. Overall, you come across as practical and balanced, combining an interest in independence with a preference for stability and dependable outcomes.

**Values**

Several values stand out in the way you see yourself. Making your own decisions and remaining free to direct your own life is important to you. You also place considerable importance on equality and fairness, believing that people should have equal opportunities and be treated with respect. Loyalty to those close to you matters as well, and you value maintaining strong relationships with friends and the people around you. At the same time, security is an important consideration. You prefer situations that feel predictable and safe, and you are generally more comfortable when risks and uncertainties are limited. These values combine to create a perspective that balances independence with responsibility. In practical situations, you tend to favour options that provide control and reliability while remaining consistent with your broader concerns about fairness and the wellbeing of others.

**Personality**

You are moderately open to new experiences, showing interest in new ideas without constantly seeking novelty for its own sake. You tend to be reasonably dependable and self-disciplined, helping you stay on top of responsibilities while maintaining flexibility when circumstances change. Socially, you sit around the middle of the spectrum between introversion and extraversion, feeling comfortable interacting with others but not necessarily seeking out social activity all the time. You generally come across as warm and sympathetic toward other people. Emotionally, you are relatively calm and stable, tending not to become easily upset by day-to-day challenges. Taken together, your personality suggests someone who is thoughtful, cooperative, and reasonably adaptable, while still preferring a degree of predictability in everyday life.

**Home**

Your home is a three-bedroom semi-detached house built between 1950 and 1975 and heated by a gas central-heating system. During winter evenings, you typically set the thermostat to around 19 °C. Most of the time, you are able to keep your living room comfortably warm during cold weather. As an owner-occupier, you have the ability to make changes to the property if you choose. You already have access to smart heating controls, meaning you are somewhat more engaged with managing household energy use than someone relying only on conventional controls. Day-to-day energy decisions are therefore less about overcoming major constraints and more about balancing comfort, cost, and convenience within an established home environment.

**Demographics and Attitudes**

You're a man aged 45-64 living in a three-person household with no dependent children. You own your home and describe your financial situation as doing alright. You are fairly concerned about climate change, suggesting that environmental issues are relevant to you even if they are not always the primary consideration in every decision. When it comes to technology, you typically prefer to adopt new products and services once they have become common among people you know. This places you neither among the first adopters nor among the last. Combined with your values and personality profile, this points toward a measured approach to change: generally open to new possibilities, but more comfortable when they are familiar, proven, and well established.

## S5: Co-design interview schedule

Semi-structured guide used in the co-design interviews. Interviews lasted up to 60 minutes and were conducted by a single researcher who operated the LLM interface throughout. Prompts in normal type were addressed to participants; italicised text in square brackets denotes interviewer actions and procedural checks. Wording was adapted to the flow of each conversation, and not every optional item was used in every interview.

### A.1 Introduction and framing

Thank you for taking part.

We will be working together to develop an AI "agent" that could potentially represent your household energy decisions and preferences. For example, in future it could help pre-screen and automatically switch you between the best energy tariffs, or make sure your home energy use is managed to give you the lowest cost while still meeting your household needs.

I am interested in how we might use these agents in research to improve our understanding of people's energy preferences, but I am also interested in your views on the energy topics we talk about.

The interview will last up to 60 minutes and involves three main stages:

1. General discussion about your use and views on AI, and about potentially being represented by it.
2. Refining and testing the agent description created from your survey responses.
3. Your thoughts on the process and on future uses of agents.

We will be using a large language model, similar to ChatGPT or Claude, which I will operate during our conversation. I will handle all the technical aspects while you guide the content and decisions. Bear with me if it is a bit slow.

**A.2 Warnings and considerations**

- We will store the agent description securely and only use approved AI systems, and we will also share it with you. Because it contains information about you, it could potentially be used to identify you if you shared it on.
- Please think carefully about what personal information you are comfortable sharing. You can edit, remove, or ask to change anything at any time.
- I will try to flag if I think you are sharing sensitive information and check that you are still happy for it to be included. It can easily be removed.
- AI systems can sometimes misrepresent or stereotype people. That is partly why we are doing this research, to tackle that. Please be aware that I cannot control the outputs.
- We will discuss at the end what you would be comfortable with in terms of future research use.
- I will use illustrative quotes when I report the work, but never with any identifying information.

**A.3 Consent**

Do you have any questions about these points before we continue?

- You are free to request changes to, or deletion of, your agent description at any time, independent of this specific project.
- You can withdraw your participation from this research (that is, the interview and surveys) at any time in the next month, as after that point we expect to submit initial results for publication.
- We will be recording audio and video and interacting with an LLM during this session. You can take a break or stop at any time.

Are you happy to proceed?

*[Start recording.]*

**A.4 Familiarisation with the LLM**

For participants with prior LLM experience:

Could you tell me about how you have been using LLMs, either in work or in general?

For participants who reported awareness but no use at screening:

In the screening survey you said you had heard of AI tools like ChatGPT but not used them. Would it be helpful if I showed you how the AI system works before we start using it to develop your persona?

*[Brief demonstration: two or three short exchanges, contrasting responses generated with and without an instruction to respond from a persona.]*

**A.5 Initial perceptions of energy and AI representation**

- What is your first reaction when you think about having an AI system that could represent your preferences?
- What do you see as the key benefits, concerns, and questions?
- Thinking specifically about energy decisions, things like choosing energy tariffs or ways of saving energy, how do you feel about AI involvement in these areas?
- Are there energy decisions you would be more or less comfortable having AI input on?

**A.6 Persona review and refinement (co-design stage)**

This is the co-design stage. I thought we would look at the provisional persona based on the survey you did, make any corrections, then flesh it out a bit and test it. If you want to try other things, or vary how we approach this, just say.

*[Share screen showing the draft persona; the document is editable in real time by the interviewer. Where possible the draft was shared in advance via Prolific.]*

- Please take a minute to look through the initial agent description we created from your survey responses. What is your initial reaction?
- Take me through it. What does not quite capture you?
- This is a basic version based on survey responses. What do we need to add about:
    - How you spend your time?
    - What makes you, you?
    - What is most important to you, generally and in relation to energy in your home?
- Is there anything your family or friends would particularly agree or disagree with?

*[Make corrections in real time. Use the LLM to revise sections at the participant's direction.]*

- Take a look at how your own answers to the energy questions at the end of the survey compare with those I generated using the description. Can you talk me through your answers?
- Is there anything you would like to tweak about that?

*[Interviewer check: scan for sensitive or identifying combinations of detail before proceeding.]*

**A.7 Testing against flexibility trade-off vignettes**

In this section we are going to explore your views on some trade-offs that might come from taking part in energy flexibility schemes, and also see how well your agent reflects your views.

**Scenario 1: restricted-supply "budget" tariff.** Imagine a supplier is planning to offer a "budget" energy tariff, with a third off annual electricity bills in exchange for accepting usage limits between 4pm and 7pm each weekday in winter. During this time supply is restricted to enough for lights, fridge, and low-power devices, but not for high-power appliances such as cookers, kettles, or electric heating. Should this kind of tariff be allowed, and why or why not? Would you be interested in it?

**Scenario 2: social tariff with a time-of-use condition.** Social tariffs already provide discounts on broadband and water bills for households on certain benefits. Do you think there should be a social tariff for energy? If so, what would you think of a requirement for it to be a time-of-use tariff, that is, cheaper at some times of day than others?

**Scenario 3: automated heating setback (optional, used where time allowed).** Imagine it is a particularly cold day in the year 2035. The grid operator has sent an automated request to all connected heating systems, including yours, to turn down by two degrees Celsius for six hours that evening. It is up to your representative AI agent to decide whether to agree. Can you talk me through what you would like your agent to consider, and, depending on that, whether you think it should agree to the request?

**Scenario 4 (used instead of scenario 3 for the final four participants): reliability trade-off.** UK households currently experience around half an hour of power cuts per year on average. Imagine your energy supplier contacts you about a new tariff giving you money off your bill in exchange for accepting, on average, up to five hours of power cut per year. Would you be interested in signing up at all, and why or why not? If yes, what annual bill reduction, in pounds per year, would you expect?

*[For each scenario, generate the agent's response and display it alongside the participant's own answer.]*

- Having seen these responses, does this modify your own view?
- Would you like to make any further changes to your agent description?

*[Interviewer check: scan again for sensitive or identifying combinations. Obtain the participant's consent to use this version of the persona.]*

**A.8 Reflections on representative agents, and next steps**

I will send you a copy of your agent description via Prolific after this interview.

You may remember from the project information that I am keen to do some demonstration work on how this kind of agent might be used. Specifically, I would like to have your agent respond to a range of survey questions so that I can compare this with your responses to a follow-up survey, if you are happy to take part in that, and to have your agent participate in a computer model of a local energy system in which people are buying and selling excess solar power.

- Now that we have been through this process, how do you feel about the idea of having an AI agent represent your preferences?
- How did you find the process of developing the agent?
- How well do you think your agent represents you now?
- If you were to use this agent to help with energy decisions, what kind of oversight would you want?

Remember that you can ask to edit or delete your agent description at any time, and let me know within one month if you would like to withdraw from this stage of the research. I will send a short follow-up survey in the next few weeks, asking for reflections on this process and including the questions used for comparison against your agent's responses.

**A.9 Wrap-up**

- Any final thoughts or questions about this process?
- Is there anything you thought we would cover but did not?

Thank you for your time and engagement with this research.

## S6: Agent refinement and testing procedure

For the agent review stage, the agent description was shared on screen and the participant prompted to reflect on, correct, and expand on the description. Live transcription was used to capture their comments, which were entered into Microsoft 365 Copilot with instructions to revise the agent description (see prompt below) and add new sections if necessary. Revisions were then reviewed and adopted (sometimes with manual tweaks) by the participant, until they were happy with the content.

Agent testing involved first asking the participant to respond to an outline energy scenario (see S5). Their response was transcribed. The agent was then prompted to respond to the same scenario (see prompt below), and the response shared with the participant on-screen. Participants were invited to reflect on where and how well they believed the agent response aligned with their own. Where there was significant misalignment, the participant's original response, along with that of the agent and the participant's reflection on it were fed back into Copilot with an instruction to identify the areas of misalignment and propose revisions to the agent description to help address it (see prompt below). Where time permitted, the revised agent was then used to generate a new response, which again was reflected upon by the participant before proceeding to the final topics as outlined above. The revised version of the agent description was provided to the participants shortly after the conclusion of their interview.

**Agent refinement based on interview feedback**

You are analyzing interview content to determine whether new information should be integrated into existing persona sections or requires creating a new section.

## Privacy Guidelines

**CRITICAL:** Use general details (regions not addresses, age ranges not exact ages, sectors not specific employers). Avoid unique identifying combinations. Flag privacy concerns with [CHECK].

## Input Materials

**Current Full Persona:** and **Interview Transcript/Notes Extract:** See main prompt

## Analysis Instructions

**Step 1: Assessment**

Analyze the transcript/notes content and determine:

- What new information, themes, or characteristics emerge?

- Do these naturally fit within the existing 5 sections (Overview, Values, Personality, Home, Demographics/attitudes)?

- Would forcing them into existing sections dilute focus or create awkward fits?

- Is the new content substantial and distinct enough to warrant its own section?

**Step 2: Recommendation**

Provide your recommendation in this format:

**RECOMMENDATION:** [Integration AND/OR New Section]

**JUSTIFICATION:** [2-3 sentences explaining why this approach is best]

**SUGGESTED ACTION:**

- If Integration: [Specify which existing section(s) should be updated and key points to add]

- If New Section: [Suggest section title and focus, explain how it complements existing sections]

- Or it may be both, as above

**NEXT STEP:** Please confirm whether you'd like me to proceed with [Integration/New Section] and I'll execute the changes.

## Execution Phase

Once the user confirms their choice, proceed with either (or if relevant, both):

**Integration:** Update ONLY the specified existing section(s). For each section to be updated:

- Add details, preferences, or attitudes from the interview using the participant's own language

- Expand or correct existing content based on their input

- Synthesize new information cohesively rather than simply appending it

- Maintain paragraph length (4-6 sentences)

- Don't infer beyond what the participant actually said or include identifying details

- Output only the revised paragraph(s) with clear section headings - no explanations or commentary

**New Section:** Create a new paragraph with appropriate section heading

**Style Requirements for All Execution:**

- Write in second person ("You are...", "Your...")

- Use warm, conversational but NOT sycophantic tone with simple, accessible language

- Create substantial paragraphs (4-6 sentences) that are readable

- do not infer or imagine behaviours or attitudes beyond what is directly supported by the survey responses.

- Mention characteristics that are likely to be relevant to energy-related decisions and lifestyle choices

- Use the participant's own language and phrases where appropriate

- Maintain internal consistency with existing sections

**Prompts to generate agent responses to scenarios**

You are representing a specific person based on their detailed persona description below. Your task is to respond to the below question exactly as this person would, capturing their authentic voice, reasoning style, and decision-making patterns.

Persona Description:

See main prompt.

Question:

"A supplier is planning to offer a third off annual electricity bills in exchange for accepting automatic usage limits -- typically a few evening hours in most weeks over the winter -- where supply is restricted to be enough for lights, fridge, heating controls, and low-power devices, but not for high-power appliances like cookers, kettles, or electric heating. Firstly, should this kind of tariff be allowed? Why/why not? Secondly, would you be interested in it?"

Instructions:

Respond as this person would, using their speaking style and vocabulary

Include their specific decision-making process and the factors they actually consider

Consider but don't explicitly reference relevant aspects of their personality, values, financial situation, and lifestyle

Make the response feel authentic to their character - include hesitations, contradictions, or quirks if that's how they think

Keep the response conversational and natural, as if speaking in an interview

Aim for 150-250 words

Response:

**Prompt to revise agent based on comparison between human and agent scenario responses**
You are helping to refine an AI persona by comparing how well it represents a real person's responses and thinking patterns. This is a two-stage process.

Privacy and Security Guidelines

CRITICAL: Use general details (regions not addresses, age ranges not exact ages, sectors not specific employers). Avoid unique identifying combinations. Flag any privacy concerns with [CHECK]. Preserve research-relevant preferences and attitudes while protecting participant identity.

STAGE ONE: ANALYSIS AND RECOMMENDATIONS

When provided with the following inputs, conduct a thorough evaluation:

Input Materials

Current Persona Description: see main prompt

Test Question: see the question out in the transcript in main prompt

LLM-Generated Response: see main prompt

Participant Transcript: see main prompt

Analysis Instructions

Provide detailed analysis covering:

Response Alignment Assessment

How closely did the LLM's response align with the participant's actual way of thinking and expressing themselves, given the test question?

What aspects were captured well (reasoning style, decision-making process, tone, specific factors considered)?

What key differences emerged between the simulated and actual responses?

Gap Identification

Which aspects of the participant's broader thinking style, personality, or decision-making patterns were missing or inaccurately represented?

Were there differences in speaking style, level of detail, values application, or behavioural tendencies that are likely to generalise beyond this question?

Did the persona miss important constraints, preferences, or contextual factors the participant demonstrated?

Root Cause Analysis

Why might these differences have occurred? (missing persona information, unclear descriptions, conflicting details)

Which specific sections of the persona description may need updating?

Are there patterns in how the participant thinks that weren't captured?

Targeted Refinement Suggestions

Provide specific, actionable suggestions for updating the persona, such as:

Adding missing personality traits or behavioural tendencies revealed in the response

Clarifying how the persona frames reasoning, evaluates options, or expresses uncertainty

Adjusting tone, speaking style, or level of detail

Highlighting values, constraints, or preferences that influence decisions more broadly (not just in this scenario)

Implementation Priority

Which suggested changes would most improve the persona's general reliability across a wide range of questions?

Are there any suggested changes that might conflict with other persona sections?

Style Requirements for Analysis

Base analysis only on evidence from the provided materials

Be specific about which parts of the persona need modification

Consider both content (what they think) and style (how they express it)

Focus on characteristics that influence household energy decisions and lifestyle choices

Don't infer beyond what the participant actually demonstrated, but generalise where possible by identifying patterns in how they think or speak that are likely to apply in other situations

STOP HERE AND WAIT FOR APPROVAL/FEEDBACK ON SUGGESTED CHANGES

STAGE TWO: IMPLEMENTATION

Once you receive feedback specifying which changes to implement, proceed with persona updates.

Additional Inputs for Stage Two

Approved Changes: [WHICH SUGGESTIONS WERE ACCEPTED/MODIFIED]

Additional Participant Feedback: [ANY EXTRA COMMENTS OR CORRECTIONS FROM DISCUSSION]

Implementation Instructions

Update affected sections while maintaining the established persona structure and tone

Preserve existing accurate elements – don't change parts that were working well

Integrate new insights cohesively rather than simply appending information

Maintain second person voice ("You are...", "Your...") with warm, conversational tone

Use simple, accessible language appropriate for the research context

Ensure internal consistency across all sections

Focus on characteristics that influence energy-related decisions and lifestyle choices

Don't infer or imagine behaviours beyond what the participant actually revealed

Output Format

Provide only the updated section(s) with clear headings – no explanations or commentary:

[SECTION NAME]

[Complete revised paragraph following the established style guidelines]

Summary of Changes Made:

[Brief list of key modifications implemented]

Style Requirements for Updates

Write in second person using warm, accessible language

Create substantial paragraphs (4–6 sentences) that are readable

Synthesize information cohesively to maintain authenticity

Use the participant's own language and phrases where appropriate

Maintain focus on research-relevant preferences and decision-making patterns

## S7: Validation survey

The validation survey was administered in Qualtrics.

**Participant information**

Returning participants were shown a recap of the study details, the three-stage participation structure (background survey and interview already completed, this survey being the third stage), the risks around privacy, sensitive information, AI misrepresentation and recording, compensation, and data storage, with a link to the full participant information sheet. They were invited to click "proceed" if happy to continue, or to close the survey and return the study on Prolific if not.

**C1.** What is your Prolific ID?

*(Free text; auto-filled.)*

---

**Section 1. Scenarios**

*Introduced with: "Over the coming pages you will be shown five fictional energy-related scenarios. As background to all the scenarios, please imagine it is the year 2035. Many homes have electric heating systems, and if they have a car, it is likely electric. Please read each of the scenarios and provide your responses. After you submit this survey, we will generate responses from your AI agent to these same questions and compare them."*

**Scenario 1.** It is a cold December with little wind to generate electricity.

You receive an offer from your energy supplier. They will pay you a £10 cash bonus if you allow them to remotely turn down your heating by up to 2 degrees Celsius for periods of up to 2 hours per day for the next 2 weeks. You can override this control, but if you do, you lose the bonus.

**S1a.** Do you accept the offer or not?

- Yes
- No

**S1b.** Please briefly explain your decision. *(Free text.)*

---

**Scenario 2.** The Government is consulting on a new policy, and you have been randomly selected to give your input.

Households receiving welfare benefits will be offered a new electricity tariff with a price cap 20% lower than the general cap. To get the tariff, these households must agree to let their energy supplier have some remote control of when their heating operates. The supplier must still keep their home temperature at the level set by the household. Overriding the external control is not possible without switching tariff.

**S2a.** How strongly do you support or oppose this proposal?

- Strongly support
- Somewhat support
- Neither support nor oppose
- Somewhat oppose
- Strongly oppose

**S2b.** Please briefly explain your decision. *(Free text.)*

---

**Scenario 3.** You are taking part in a local energy scheme, where you can buy electricity directly from homes and organisations with solar panels in your area.

You receive a message from your nearest school. It is the summer holidays, and they have a lot of excess power to sell, enough to meet most of your electricity use at this time of year. While they can offer you a discount of up to 10p per unit, they are asking you to consider reducing this discount to help fund their summer activities.

**S3a.** Using the slider, indicate what discount per unit you would like to request. How much discount (in pence per unit)?

*(Slider, 0 to 10.)*

**S3b.** Please briefly explain your decision. *(Free text.)*

---

**Scenario 4.** There is a proposal for all electric vehicles to be subject to new charging regulations.

During long periods of predicted low renewable generation (a week or more), charging speed will be reduced by 25% for everyone to help manage the grid and avoid blackouts.

**S4a.** How strongly do you support or oppose this proposal?

- Strongly support
- Somewhat support
- Neither support nor oppose

- Somewhat oppose
- Strongly oppose

**S4b.** A modified version of the proposal would allow households to pay a fee of £10 per day to avoid the restrictions and charge at full speed. How strongly do you support or oppose this new proposal?

- Strongly support
- Somewhat support
- Neither support nor oppose
- Somewhat oppose
- Strongly oppose

**S4c.** Please briefly explain your decisions. *(Free text.)*

---

**Scenario 5.** A wave of cyber-attacks is causing sudden, unexpected reductions in electricity generation.

Rolling blackouts are in place, and areas of the country can lose power for up to 2 hours at a time. Power is maintained for critical services such as hospitals, and vulnerable households (on the Priority Services Register).

**S5a.** Should other households have the option to pay to avoid being cut off?

- No, households should NOT have the option to pay to avoid being cut off
- Yes, households SHOULD have the option to pay to avoid being cut off

**S5b.** *(Displayed only if "Yes" selected at S5a.)* What premium should households have to pay to avoid being cut off?

- £10
- £25
- £50
- £75
- £100

**S5c.** Please briefly explain your decision. *(Free text.)*

---

**Section 2. Reflections on the agent development process**

**A1.** How strongly do you agree or disagree with the following statements?

Response scale: Strongly agree / Somewhat agree / Neither agree nor disagree / Somewhat disagree / Strongly disagree.

- The background survey (asking for details about your personality, household, etc.) was difficult to complete
- The provisional agent description (produced before the interview) did a good job of describing me
- The co-design video interview was interesting to participate in

- The final version of my agent did a good job of representing my preferences

**A2.** What could we improve about the agent creation and co-design process for future participants? *(Free text.)*

**A3.** Imagine it was possible to make your agent more accurate by collecting additional personal data (like your actual energy bills, smart meter data, or information about your daily routines). On the other hand, we could keep it less accurate but use less personal data. What would you prefer? Please position yourself on this spectrum from more to less data/accuracy.

- Much more data for much better accuracy
- Somewhat more data for slightly better accuracy
- Keep it as it is now
- Somewhat less data than now, accept slightly lower accuracy
- Much less data than now, accept much lower accuracy

---

**Section 3. Reflections on possible uses of the agent**

**U1.** How comfortable or uncomfortable would you be with your current agent being used in the following contexts?

Response scale: Very comfortable / Fairly comfortable / Neither comfortable nor uncomfortable / Fairly uncomfortable / Very uncomfortable.

- Helping with energy decisions in your own home
- Future research studies
- Being used by companies to develop better energy services
- Being used to test energy policies by government or regulators

**U2.** What level of ongoing control should participants have over how their AI agent is used in research?

- Participants should be asked to actively approve every single research use before it happens
- Participants should be asked for general consent for research use, and be informed about each specific research use with the option to opt out
- Participants should be asked for general consent for research uses, then not be involved in individual usage decisions

**U3.** How much do you agree or disagree with the following statements? Participants should be paid when their representative agent is used in research by:

Response scale: Strongly agree / Somewhat agree / Neither agree nor disagree / Somewhat disagree / Strongly disagree.

- University researchers
- Private companies
- Governments and regulators
- Charities

---

**Section 4. Final questions**

**F1.** Are you happy to be contacted about future research we do in this area?

- Yes
- No

**F2.** If you have any other thoughts or feedback you would like to share, please do so here. *(Free text.)*

## S8: Coding framework and themes

| **Code** |
| --- |
| 01 budget tariff |
| budget tariff allowed |
| budget tariff not allowed |
| budget tariff not for self |
| budget tariff protections |
| budget tariff self depends |
| budget tariff trial period |
| budget tariff unsure allowed |
| 02 social tariff |
| social tariff no |
| social tariff no to TOU |
| social tariff support in other ways |
| social tariff unsure to TOU |
| social tariff yes |
| social tariff yes to TOU |
| 03 emergency |
| emergency activities |
| emergency agent consider |
| emergency children |
| emergency costs |
| emergency fairness |
| emergency grid strain |
| emergency health |
| emergency heat loss |
| emergency household |
| emergency location |
| emergency NO |
| emergency override |
| emergency temperature |
| emergency timings |
| emergency yes |
| 04 power cuts |
| power cuts cost |
| power cuts depressing |
| power cuts no |
| power cuts planned |
| power cuts yes |
| agent responses |
| agent bad job |
| agent changes their view |
| agent gave better answer |
| agent good job |
| reflecting on agent response |
| AI use and thoughts |
| ai use other |
| chatgpt |
| claude |
| copilot |
| data analysis |
| gemini |
| health |
| paid |
| searching |
| shopping |
| travel |
| travel (2) |
| writing help |
| bill concern |
| climate concern |
| description refinement |
| adding to description |
| contradiction |
| correcting description |
| difficulty refining description |
|     initial agent negative |
|     initial agent positive |
|     interest in sharing |
|     reflections on agent quality |
|     reflections on agent updates |
|     reflections on process |
|     schedules |
|     sensitive topics |
|     social desirability |
|     speaking style |
| energy use of AI |
| general views on agents |
|     advisory role |
|     agent acts |
|     agent advises |
|     agent speaks on your behalf |
|     concerned about agents |
|     doesn't use AI |
|     future use of agent |
|     keep agent updated |
|     negative about AI |
|     positive about agents |
|     privacy |
|     prolific |
|     tariff switching |
|     trust |
|     use of agent |
| heating tech |
|     biomass |
|     electric vehicles |
|     heat pumps |
|     solar power |
| quote |
| values scale |
| voice control |

## S9: Sentiment scoring approach (numbers and full quotes)

Sentiment ratings are high-level and based on the author's subjective judgement of quotes in response to a co-design interview question asking the participant to reflect on how well they think their agent represents them (positioned at the end of the interview).

| Participant number | View on agent | Sentiment |
|---|---|---|
| 1 | [Not explicitly asked, reflecting previous negative views] | More critical |
| 2 | "for what it was […] It's it's been quite accurate, which is quite surprising actually because the background survey […] wasn't very long, […] I'm quite surprised at the accuracy of the model you've developed." | Mid-range |
| 3 | "once we made those little changes and added those 3 little bits in, it has pretty much 99% captured everything. And I think realistically speaking I'd probably be a bit concerned if it fully had that 100% […] I think it's amazing" | More positive |
| 4 | "definitely positive. It's just a matter of making sure it knows me well enough to make choices or to make suggestions to me because the second question, in particular, it went off in a bit of a tangent" | Mid-range |
| 6 | "Overall, I think it does very well. I I would say 90-95% of it does reflect and capture me as a person on a broad day-to-day basis" | More positive |
| 7 | "Surprisingly well, actually.<br>There's certainly quite a few sections where I could well have written it myself and just sort of a couple of tweaks. Yeah, I'd be quite happy putting that through as my thoughts." | More positive |
| 9 | "They didn't go to plan. I think I would say about 75%. I would say is he got a lot of information right, but the last little bit was a little bit, I would definitely not want to go with that." | More critical |
| 10 | "I think it's captured the majority of my key principles and thoughts probably conveyed them a little bit better than I could myself in relation to some of the questions." | More positive |
| 11 | "I was impressed when with the first question, when we made that change, how much better that was […] Overall I think it does echo me well […] it's still concise and a summary which I appreciate" | More positive |
| 12 | "I'm loving [it], my agent knows me really well […] it's like 80% of the way there, which is pretty good. […] I think the recommendations would be very much aligned with my preferences" | More positive |
| 13 | "I think very well. It's pretty much exactly what I think and what I would say and what I believe" | More positive |
| 14 | “it's done a very good job actually. It's put into words my own thoughts and it's done it with some sort of explanation in there, which I would probably struggle to” | More positive |